\documentclass[aps,twocolumn,groupedaddress,showpacs,amssymb,prl,floatfix,preprintnumbers]{revtex4}
\usepackage{graphicx,color}
\usepackage{amssymb}

\newcommand{\slrr}      {$T_1^{-1}$}

\newcommand{\cecoin}    {CeCoIn$_5$}
\newcommand{\cemin}    {CeMIn$_5$}

\newcommand{\tn}     {$T_{\rm N}$}

\newcommand{\slrrtext}  {spin-lattice-relaxation rate}

\newcommand{\ceptin}     {CePt$_2$In$_7$}

\begin{document}

\thispagestyle{myheadings}

\title{Commensurate Antiferromagnetism in CePt$_2$In$_7$, a Nearly Two-Dimensional Heavy Fermion System}

\author
{N. apRoberts-Warren, A. P. Dioguardi, A. C. Shockley, C. H. Lin, J. Crocker, P. Klavins, and  N. J. Curro \email{curro@physics.ucdavis.edu}}

\affiliation{Department of Physics, University of California, Davis, CA 95616}


\date{\today}

\begin{abstract}
CePt$_2$In$_7$ is a new heavy fermion system with a structure similar to the CeMIn$_5$ system but with greater two-dimensional character.  We report the synthesis and characterization of phase pure polycrystalline material, which reveals an antiferromagnetic transition at $T_N = 5.2$ K.  Nuclear Quadrupolar Resonance (NQR) studies indicate that the antiferromagnetism is commensurate and exhibits strong antiferromagnetic fluctuations in the paramagnetic state.

\end{abstract}


\maketitle

In the quest for new high temperature superconductors, two guiding principles have emerged in recent years:  (1) reduced dimensionality and (2) proximity to antiferromagnetism.   Materials which satisfy these two criteria are more likely to lead to unconventional superconductivity upon doping or application of pressure \cite{MonthouxPinesReview,uedareview}. Several years ago, this design principle lead to the discovery of the \cemin\ (M = Co, Rh, Ir) heavy fermion system, which has revealed a wealth of information about the subtleties and interplay between antiferromagnetism and superconductivity \cite{jdtreview}.  The basic structural elements of these materials are Ce-In planes, which give rise to a two dimensional character of the electronic behavior and a ten-fold increase in the superconducting transition temperature from the cubic CeIn$_3$ system.    Here we report data in a new variant, \ceptin, in which the spacing between Ce-In planes is increased by 43\% \cite{kurenbaeva}.  Recently, Bauer and coworkers discovered that \ceptin\ is antiferromagnetic at ambient pressure, and becomes superconducting under pressure \cite{Bauer127Report}.   Our microscopic NQR measurements reveal that this system is antiferromagnetic below $T_N = 5.2$ K, with staggered antiferromagnetic ordering in-plane, and collinear antiferromagnetic correlations between adjacent planes. Measurements of the \slrrtext\ reveal enhanced fluctuations at temperatures up to nearly 4\tn.  This behavior suggests that the antiferromagnetic fluctuations in \ceptin\ are more two dimensional than in the \cemin\ system, where the spin fluctuations are present only up to at most 1.25\tn.

\begin{figure}
\includegraphics[width=1.0\linewidth]{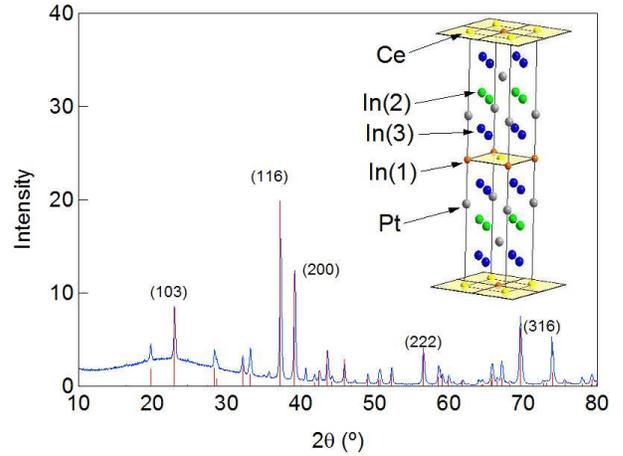}
\caption{\label{fig:xray} (color online) X-ray powder diffraction and structure. Vertical red lines are theoretical peak positions.}
\vspace{-0.15in}
\end{figure}

Polycrystalline samples of \ceptin\ were synthesized by first arc-melting stoichiometric quantities of Ce, Pt, and In under Ar atmosphere. The reacted ingot was wrapped in Ta foil, sealed under vacuum in a quartz ampoule and annealed for one week.  Powder x-ray diffraction revealed phase pure material that crystallized in the space group $I4/mmm$ with lattice constants $a=4.602$ {\AA} and  $c=21.601$ {\AA} (Fig. \ref{fig:xray}).  The planar Ce-Ce spacing in \ceptin,  $a$, is reduced by 8\% from that of cubic CeIn$_3$, and is 1\% smaller than in CeRhIn$_5$.  The interplanar distance along the $c$ direction is 10.8 {\AA} in \ceptin, 43\% larger than in CeRhIn$_5$. Two  important structural differences between these two compounds are the alternating Ce-In(1) planes along the $c$ direction that are displaced by half a lattice constant along the [110] direction, and a third set of In sites located halfway between the Ce-In(1) planes that serve to expand the lattice along $c$ direction.

\begin{figure}
\includegraphics[width=1.0\linewidth]{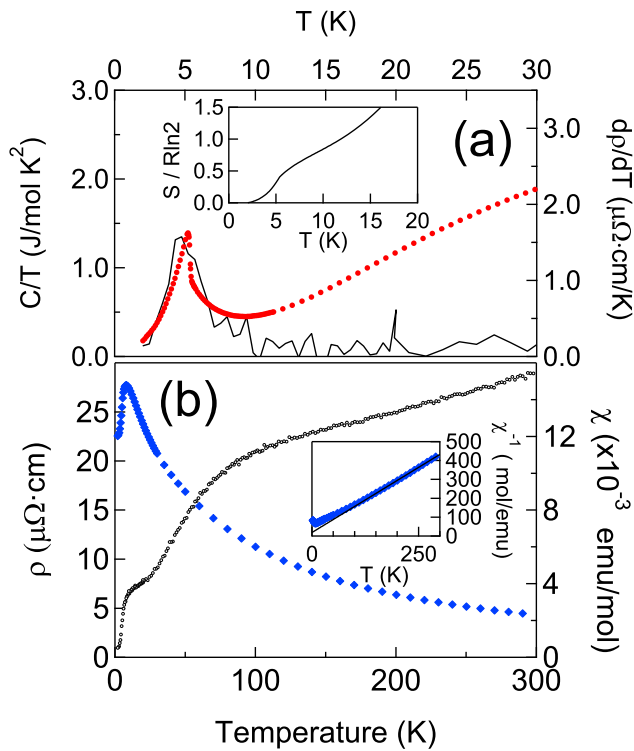}
\caption{\label{fig:bulkdata} (color online) Upper panel: Total specific heat ($C$) divided by temperature ($\bullet$, left axis) and $d\rho/dT$ (solid line, right axis) versus temperature. The inset shows the total entropy obtained by integrating $C/T$ versus temperature. Lower panel:  resistivity ($\circ$) and susceptibility ($\blacklozenge$) versus temperature; the inset shows the inverse of the susceptibility versus temperature, as well as a linear fit.}
\vspace{-0.15in}
\end{figure}

Figure \ref{fig:bulkdata} summarizes the bulk thermodynamic properties of polycrystalline \ceptin, which are in agreement with those first reported in \cite{Bauer127Report}.  The specific heat divided by temperature, $C/T$, exhibits a minimum at 9.1 K with a value of 450 mJ/mol K$^2$ and a peak at 5.2 K associated with an antiferromagnetic transition. The large value of $C/T\sim m^*$ indicates a substantial enhancement of the effective mass, $m^*$, arising from the interactions of the conduction electrons with the magnetic Ce ions \cite{zachreview}.  For comparison, $C/T\sim 290-750$ mJ/mol K$^2$ in the \cemin\ compounds \cite{CeCoIn5discovery}. The inset shows the entropy obtained by integrating the total specific heat as a function of temperature, and reveals a value of $\sim 36$\% $R\log 2$ at \tn.  Although  the specific heat and entropy include contributions both from the magnetic 4f electrons and the lattice, the large value of $C/T$ and the small value of $S(T_N)$  suggest that the magnetic moment in the ordered state is reduced by Kondo interactions, similar to CeRhIn$_5$ \cite{jdtreview}.  The magnetic susceptibility, $\chi(T)$, of the polycrystalline powder shows Curie-Weiss behavior for $T\gtrsim 150$ K, with an effective moment of 2.41 $\mu_B$, consistent with Ce$^{3+}$ ($p = 2.54$), and a Curie-Weiss temperature of 13.5 K. $\chi$ exhibits a maximum at 8K, and an inflection at \tn.  The resistivity, $\rho(T)$, shows metallic behavior with inflections at 100 K and 25K, and a sharp suppression below 8 K with a residual resistivity $\rho_0 = 1.9 \mu\Omega\cdot$cm.  The origin of these two higher temperature anomalies is unknown, but may be related to either a crystalline electric field excitation and/or the onset of coherence of the Kondo lattice.  Indeed, the inflection point in $\rho(T)$ at 100 K in \ceptin\ is qualitatively similar to one observed in  CeRhIn$_5$ at roughly 40 K that has been interpreted as the coherence temperature \cite{YangPinesNature}.  The temperature dependence of $d\rho/dT$ scales roughly with $C/T$ (Fig. \ref{fig:bulkdata}(a)) with a peak at \tn, indicating the presence of antiferromagnetic fluctuations \cite{FisherLanger}. The sharp downturn in $\rho(T)$ and $\chi(T)$ at 8 K signals the onset of these fluctuations.

\begin{figure}
\includegraphics[width=1.0\linewidth]{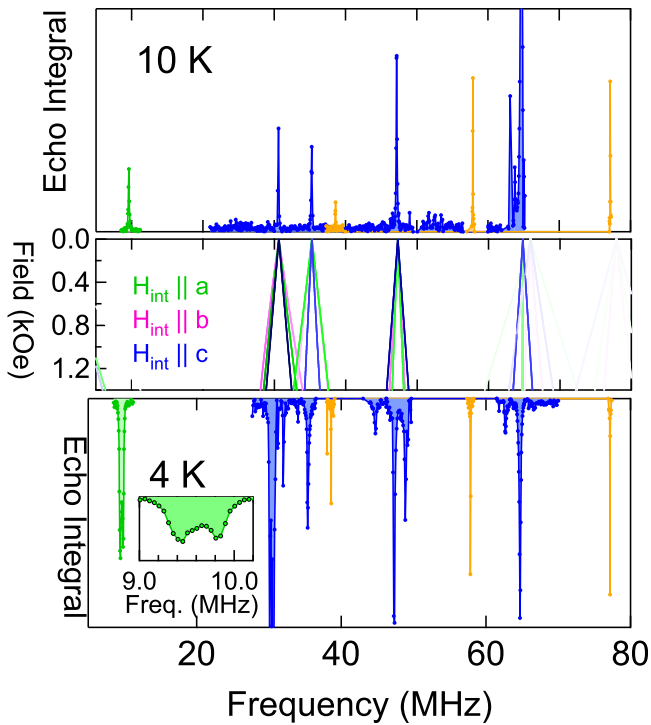}
\caption{\label{fig:spectra} (color online) NQR spectra above \tn\ (top panel), below \tn\ (bottom panel), and (middle panel) the field dependence of the resonant frequencies of the In(3) site for internal fields along the $c$ axis (blue), and in the $ab$ plane parallel $|| a$, green) and perpendicular ($|| b$, magenta) to the normal on the unit cell face \cite{CurroAnomalousShift}.  Color intensity is proportional to the matrix element for the particular transition.  Colors for the In(1), In(2) and In(3) resonances match those in Fig. \ref{fig:xray}.  The inset in the bottom panel shows the In(2) spectrum at 4 K.}
\vspace{-0.15in}
\end{figure}

In order to investigate the antiferromagnetic state in more detail, we have conducted detailed NMR and NQR measurements in the paramagnetic and antiferromagnetic states. $^{115}$In ($I=9/2$) is an excellent nucleus for NMR/NQR, and \ceptin\ has three crystallographically distinct In sites (see Fig. \ref{fig:xray}).  The nuclear spin Hamiltonian is given by:
$\mathcal{H}_Q = \frac{h\nu_{z}}{6}[(3\hat{I}_z^2 - \hat{I}^2) +
\eta(\hat{I}_x^2-\hat{I}_y^2)]$,
where $h\nu_{z} = 3eQV_{zz}/2I(2I-1)$, $Q$ is the quadrupolar moment, $\eta
=(|V_{xx}|-|V_{yy}|)/|V_{zz}|$,
and $V_{\alpha\beta}$ are the components of the electric field gradient (EFG) tensor \cite{NQR115electronic}. Fig. \ref{fig:spectra} shows the NQR spectrum at 10 K obtained by acquiring spin echoes as a function of frequency. The spectrum reveals three sets of resonances with distinct EFGs: site A with $\nu_z = 19.290(1)$ MHz and $\eta = 0$, site B with $\nu_z = 16.47(2)$ MHz and $\eta = 0.47(1)$, and site C with $\nu_z\approx 2.5$ MHz and $\eta\approx 0.4$.  Site C was identified by acquiring spectra in an applied magnetic field (not shown) and fitting the resulting powder pattern. Site A must be the In(1) site because it has axial symmetry.  However, it is surprising that the EFG at the In(1) site in \ceptin\ is about three times larger than in the \cemin\ series \cite{NQR115electronic}.  This enhancement may reflect a greater 2D character of the electronic structure.  The EFG at site B is similar to that of the In(2) in the \cemin\ materials. In fact, the response of these sites in the antiferromagnetic state (discussed below) suggests proximity to the Ce planes as in \cecoin\ \cite{CurroJLTP}.  We therefore assign site B to the In(3) and site C to In(2). (Note that the symmetry of the In(3) in \ceptin\ matches that of the In(2) in \cemin.)  The $z$-direction of the EFG tensor is generally defined as the principal axis with the largest eigenvalue.  In the case of In(1) the z-direction of the EFG points along the crystal $c$ axis.  In order to identify the $z$ axis of the In(2) and In(3) sites, single crystals are necessary to investigate the field-dependence of the resonances. However it is known that in the \cemin\ materials the $z$ axis of the In(2) sites lies along the normal to the unit cell faces. We therefore assume that the $z$ axis of the EFG tensor also lies along the unit cell face normals in \ceptin\ based on the similarity between the symmetries of the In(2) in the \cemin\ and the In(2,3) in the \ceptin.

Below \tn, the spectra of the In(1) broaden slightly (by $\sim$ 25 kHz), the In(2) splits equally (see inset of lower panel of Fig. \ref{fig:spectra}), and the In(3) site develops multiple peaks due to the presence of a static internal magnetic field, $\mathbf{H}_{\rm int}$ (Fig. \ref{fig:spectra}).  The middle panel of Fig. \ref{fig:spectra} shows the splitting of these NQR resonances as a function of internal magnetic field (calculated by diagonalizing $\mathcal{H}_{Q} + \gamma \hbar\hat{\mathbf{I}}\cdot\mathbf{H}_{\rm int}$, where $\gamma$ is the gyromagnetic ratio). The spectra indicate that approximately one half of the In(3) experience an internal field $H_{\rm int} = 1.4$ kOe along the $c$ axis, and that all of the In(2) experience an internal field of 0.4 kOe along the $c$ direction. In the ordered state of CeRhIn$_5$ the internal field at the In(2) also lies along the $c$ direction and is of the same magnitude, which suggests a similar magnitude of ordered moment and magnetic structure for \ceptin\ \cite{CurroCeRhIn5}.   CeRhIn$_5$ exhibits a spiral magnetic structure with moments $\mathbf{S}_0$ in the $ab$ plane and ordering wavevector $\mathbf{Q} = (\frac{\pi}{a},\frac{\pi}{a},0.297\frac{\pi}{c})$, which gives rise to a broad double-peak spectrum at the In(2) site \cite{CurroCeRhIn5,baoCeRhIn5INS}.   However in the case of \ceptin,  we observe a distinct set of resonance lines in the antiferromagnetic state rather than a broad double peak structure characteristic of incommensurate magnetism.

\begin{figure}
\includegraphics[width=1.0\linewidth]{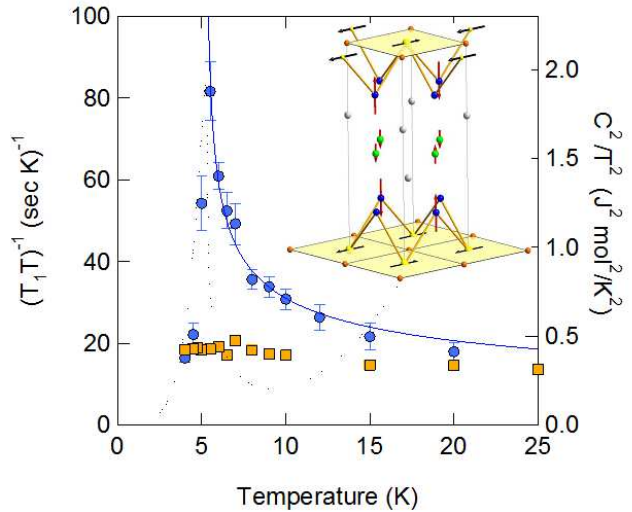}
\caption{\label{fig:T1Tinv} (color online) $(T_1T)^{-1}$ versus $T$ measured at the In(1) $3\nu_Q$ transition ($\blacksquare$) and In(3) 35 MHz transition ($\bullet$).  The In(1) data has been scaled by a factor of five.  The dotted line is the square of the specific heat divided by temperature, and the solid line is a fit to $(T_1T)^{-1} = A/(T-T_N)^{\alpha}$, where $A = 55(1)$ sec$^{-1}$ K$^{-1}$ and $\alpha = 0.36(1)$.  INSET: The ordered Ce moments (black arrows) and hyperfine fields (red arrows) consistent with the spectral observations, as discussed in the text.}
\vspace{-0.15in}
\end{figure}

Our observations of the internal hyperfine fields place important constraints on possible magnetic structures in this material: (i) $\mathbf{H}_{\rm int}(1) = 0$ at the In(1) site, (ii) $\mathbf{H}_{\rm int}(2) ~ || ~ c$, and (iii)  two sets of fields $\mathbf{H}_{\rm int}(3) = 0$ or  $\mathbf{H}_{\rm int}(3) ~|| ~\hat{c}$  at the In(3) site.  In order to deduce a possible magnetic structure, $\mathbf{S}(\mathbf{r})$, we assume a hyperfine tensor, $\mathbb{B}$, for the In(2) and In(3) sites that has the same symmetry as that for the In(2) in the \cemin\ materials, with $B_{ab} = 0$, and $B_{aa}$, $B_{bb}$, $B_{cc}$ and $B_{ac}$ nonzero \cite{CurroJLTP,CurroM2Sproceeds}.  We find that for $\mathbf{Q} = (\frac{\pi}{a},\frac{\pi}{a})$ in the basal plane, $\mathbf{S}_0~ ||~ [100]$ with collinear spins in neighboring planes, the hyperfine field at the In(3) site is given by:  $\mathbf{H}_{\rm int}(3) = 0$ or $\pm 2B_{ac}S_0\hat{\mathbf{c}}$  (see Fig. \ref{fig:T1Tinv} inset). All of the In(2) sites experience a hyperfine field either parallel or antiparallel to the $c$ direction. This particular magnetic structure is similar to that observed in La$_2$CuO$_4$, but other structures may be possible \cite{aharony}.  Without independent measurements of $B_{ac}$ we are unable to estimate the magnitude of $S_0$.  If we use the values appropriate for \cecoin, we find $S_0\approx 0.08$ $\mu_B$ at 4 K, but expect this to grow at lower temperatures \cite{CurroM2Sproceeds}. At the In(1) site, the internal field vanishes because the site is symmetrically located in the plane of the Ce atoms, similar to the case of field-induced magnetism in \cecoin\ \cite{CurroCeCoIn5FFLO}. We note that $H_{\rm int}(1)\neq 0$ in CeRhIn$_5$ because of a slight asymmetry in the hyperfine tensor in that compound, which seems to be absent in the case of \ceptin\ \cite{CurroCeRhIn5}.

In order to investigate the dynamics of the phase transition, we have measured the \slrrtext, \slrr, as a function of temperature at both the In(1) and In(3) sites as shown in Fig. \ref{fig:T1Tinv}.  For the In(1), the relaxation was measured at the 3$\nu_Q$ transition by inverting the magnetization and fitting the recovery to the standard expression for a spin $I=9/2$ nucleus.  The same expression was used to fit the relaxation of the In(3) transition at roughly 35 MHz.  We caution that since the In(3) has $\eta>0$ and hence $[\hat{I}_z,\mathcal{H}_{Q}]\neq 0$, the magnetization recovery acquires a complicated dependence on the EFG parameters and \slrr\ depends on spin fluctuations in all three directions \cite{CPSbook}.  The \slrr\ values that we extract provide a reasonable description of the temperature dependence but may not accurately reflect the absolute value of the spin fluctuations (see Eq. 1 below).

There are dramatic differences between the temperature dependences of \slrr\ at the In(1) and the In(3) sites. The \slrrtext\ is driven by fluctuations of the hyperfine fields at the Larmor frequency and is related to the dynamical spin susceptibility by the relation:
\begin{equation}
\frac{1}{T_1T} = \gamma^2k_B\lim_{\omega\rightarrow 0}\sum_{\mathbf{q},\beta}F^2_{\beta}(\mathbf{q})\frac{\chi_{\beta}''(\mathbf{q},\omega)}{\hbar\omega},
\end{equation}
where the form factor $F(\mathbf{q}$) is the Fourier transform of the hyperfine coupling and $\chi_{\beta}''(\mathbf{q},\omega)$ is the dynamical susceptibility \cite{Moriya1974,currohyperfine}.  For the In(1) site the form factor vanishes for $\mathbf{q}\sim\mathbf{Q}$, hence \slrr\ at the In(1) site is insensitive to the critical dynamics associated with the phase transition. This is reflected both in the absence of any significant static internal fields at this site below \tn, as well as the lack of any enhancement in $(T_1T)^{-1}$ above \tn.  On the other hand, the In(3) form factor does not vanish at $\mathbf{Q}$ and hence $(T_1T)^{-1}$ probes the critical slowing down of the antiferromagnetic spin fluctuations.    This interpretation is supported by the fact that although both quantities diverge at \tn, $(T_1T)^{-1}$ does not scale with $(C/T)^2$, as would be expected if the density of states, $N(0)$, were enhanced by an increasing $m^*$.  Rather, \slrr\ is dominated by slow spin fluctuations at the In(3) site, consistent with the observation of large static internal fields below \tn.

 We find that the temperature dependence is well fit by the expression $(T_1T)^{-1}=A(T-T_N)^{-\nu}$, where $\nu = 0.36(1)$.   A similar divergence was found in CeRhIn$_5$ with $\nu=0.30$ \cite{curroPRL}.  In both cases this exponent is suppressed from the mean-field value $\nu=1/2$ in three dimensions.  This suppression may suggest that the critical point $T_N(H=0)$ is in fact a multicritical point, and that other field-induced phase transitions  may emerge in applied fields, $H$ \cite{CeRhIn5fieldinducedphases}.  In contrast to CeRhIn$_5$, the critical fluctuations in \ceptin\ appear to extend up to 4\tn.  This result suggests that antiferromagnetic correlations develop near 20 K, but weak interplanar coupling suppresses the long range order \cite{PinakiRajiv}.

In conclusion, we have grown and characterized polycrystalline samples of \ceptin, and found that this nearly two-dimensional heavy fermion system with $C/T\sim 450$ mJ/mol K$^2$ above \tn\ exhibits commensurate antiferromagnetism.  This material represents a unique opportunity to investigate Kondo lattice physics in two dimensions with a weak interplanar coupling and is an ideal candidate to exhibit a quantum phase transition and superconductivity under pressure \cite{Bauer127Report}. This novel system promises to reveal new information about the basic physics unconventional superconductivity and quantum criticality in the Kondo lattice \cite{doniach,CurroPuCoGa5}.


This work was inspired by the announcement of the discovery of this material by the LANL group on strongly correlated electron materials at the ICM and M2S conferences in 2009.  We thank  D. Pines, M. Matsumoto, S. Savrasov, R. Singh, and R. Scalettar  for stimulating discussions. This research was sponsored by the National Nuclear Security Administration under the Stewardship Science Academic Alliances program
through DOE Research Grant \#DOE DE-FG52-09NA29464.


\end{document}